\documentclass[twocolumn,floats,floatfix,showpacs,amssymb,prd,superscriptaddress,nofootinbib]{revtex4-1}
\usepackage{graphicx, epsfig, amssymb}
\usepackage{epstopdf}
\usepackage{amsmath, amsfonts}
\usepackage{mathrsfs}
\usepackage{bm}
\usepackage[linktocpage]{hyperref}
\usepackage[caption=false]{subfig}
\usepackage[usenames]{color}

\begin{document}

\title{\large Comments on initial conditions for the Abraham-Lorentz(-Dirac) equation}

\author{Ofek Birnholtz} \email{ofek.birnholtz@mail.huji.ac.il}
\affiliation{Racah Institute of Physics, Hebrew University, Jerusalem 91904, Israel.}

\date{\today}

\def\blue{\color{blue}}		

\newcommand{\pao}{\bf\color{red}}
\newcommand{\eb}[1]{\textcolor{blue}{[{\it\textbf{EB: #1}}]} }

\def\be{\begin{equation}}
\def\ee{\end{equation}}
\def\bea{\begin{eqnarray}}
\def\eea{\end{eqnarray}}
\def\beq{\begin{eqnarray}}
\def\eeq{\end{eqnarray}}
\def\non{\nonumber}
\def\nn{\nonumber}
\def\L{\mathcal{L}}
\def\p{\partial}
\def\half{{\textstyle{1\over2}}}
\def\ss{\scriptscriptstyle}
\renewcommand{\vec}[1]{\boldsymbol{#1}}
\newcommand{\us}[1]{\textcolor{green}{[{\it\textbf{US: #1}}]} }
\newcommand{\vc}[1]{\textcolor{blue}{[{\it\textbf{VC: #1}}]} }

\def\IInt{\int\!\!\!\!\!\int}
\def\IIInt{\int\!\!\!\!\!\int\!\!\!\!\!\int}

\def\a{\alpha}
\def\bt{\beta}
\def\d{\delta}
\def\D{\Delta}
\def\eps{\epsilon}
\def\veps{\varepsilon}
\def\Gm{\Gamma}
\def\gm{\gamma}
\def\k{\kappa}
\def\l{\lambda}
\def\L{\Lambda}
\def\o{\omega}
\def\O{\Omega}
\def\vph{\varphi}
\def\vpi{\varpi}
\def\s{\sigma}
\def\S{\Sigma}
\def\vsgm{\varsigma}
\def\t{\theta}
\def\T{\Theta}
\def\vt{\vartheta}
\def\z{\zeta}
\def\X{\times}


\newcommand{\refeq}[1]{eq.~(\ref{#1})}
\newcommand{\refig}[1]{fig.~(\ref{#1})}
\newcommand{\Refeq}[1]{Eq.~(\ref{#1})}
\newcommand{\Refig}[1]{Fig.~(\ref{#1})}

\newcommand{\h}[1]{{\hat #1}} \newcommand{\tl}[1]{{\tilde #1}}

\def\al{\alpha} \def\bt{\beta}
\def\kp{\kappa}
\def\gm{\gamma} \def\dl{\delta} \def\eps{\epsilon}
\def\teps{\tilde{\epsilon}}
\def\hal{{\hat \alpha}} \def\hbt{{\hat \beta}}
\def\hgm{{\hat \gamma}} \def\hdl{{\hat \delta}}
\def\trho{{\tilde \rho}} \def\hrho{{\hat \rho}}
\def\hmu{{\hat \mu}} \def\sig{\sigma}
\def\lam{\lambda} \def\hxi{{\hat \xi}}
\def\w{\omega}
\def\Om{\Omega}

\def\({\left(} \def\){\right)}
\def\[{\left[} \def\]{\right]}
\def\d{\partial}
\def\sumint{\int\!\!\!\!\!\!\!\sum}

\def\scrim{\mathscr{I}^-}

\newcommand{\sect}[1]{\setcounter{equation}{0}\section{#1}}
\renewcommand{\theequation}{\thesection.\arabic{equation}}

\newcommand{\tab}{\hspace*{3em}}

\setcounter{section}{0}
\setcounter{equation}{0}

\newcommand{\VERBOSE}[1]{#1}



\begin{abstract}

An accelerating electric charge coupled to its own electromagnetic (EM) field both emits radiation and experiences the radiation's reaction as a (self-)force.
Considering the system from an Effective Field Theory perspective, and using the physical initial conditions of no incoming radiation can help resolve many of the problems associated with the often considered ``notorious" Abraham-Lorentz / Abraham-Lorentz-Dirac equations.\end{abstract}

\maketitle


\sect{Background \& Motivation}					%
\label{Intro}								%
Viewed as a single-particle equation, the century old equation of Abraham-Lorentz (AL) \cite{Abraham,Lorentz1,Lorentz2} for the self-force on an accelerating electric charge has been notorious for involving many physically and mathematically unwanted behaviours \cite{Dirac, ALD-Jackson, ALD-Rohrlich, Rohrlich2, Plass, Driver, WheelerFeynman, Murdock, Spohn, HerasDebate, Eliezer, LandauLifshitz, MoPapas, Caldirola, FordO'Connell, deOcaCaboBizet, Yaghjian, Medina, Sokolov, Hadad, ModifiedWithNumerics, Laue, Fermi, Rohrlich3, SchwingerEM, Steane, Rohrlich1, Parrott, Kosyakov, Knoll}.
The AL equation,
\be
m\ddot{\bf x} = \frac{2}{3} \frac{q^2}{c^3} \dddot{\bf x} + F_{ext}({\bf x}, \dot{\bf x}),
\label{AL}
\ee
as well as Dirac's \cite{Dirac} relativistic generalization (the Abraham-Lorentz-Dirac (ALD) equation
\footnote
{Relativistically, ${\bf x}$ becomes $x^{\mu}$, time derivatives are taken w.r.t proper time, and the jerk $\dddot{\bf x}$ is to be replaced by 
$\dddot{x}^{\mu} - \dot{x}^{\mu} \dot{x}_{\mu} \dddot{x}^{\nu}$. Note all equations are given in cgs units.
}),
involve the time derivative of the acceleration (``jerk"), and are thus $3^{rd}$ order ODE, unusual in mechanics.
The most troubling of the problems associated with them include
\begin{itemize}
\item the requirement of a $3^{rd}$ initial condition
\item causality violations due to pre-acceleration
\footnote
{For example Dirac's own remarks,``it is possible for a signal to be transmitted faster than light through the interior of an electron.
The finite size of the electron now reappears in a new sense, the interior of the electron being a region of failure, not of the field equations of electromagnetic theory, but of some of the elementary properties of space-time" \cite{Dirac}, see also \cite{HerasDebate}.
},
\item runaway/unstable solutions
\footnote
{e.g. Medina's  ``the Lorentz-Abraham-Dirac formula for the radiation reaction of a point charge predicts unphysical motions that run away or violate causality" \cite{Medina}.
},
\item disagreement with the Larmor formula for energy output of a constantly accelerating particle \cite{FordO'Connell}.
\end{itemize}
Some considerations of the origin of the ALD force in the microscopic structure of the so-called point-particle have even lead to problems such as the ``$\frac{4}{3}$ problem" \cite{Laue,Fermi,Rohrlich3,SchwingerEM}, negative mass \cite{Steane,Kosyakov}, and even suggestions that the fundamentals of EM of point particles be altogether replaced \cite{WheelerFeynman, Knoll}.
Most attempts for solving these problems have suggested modifying the equation itself \cite{Eliezer, LandauLifshitz, MoPapas, Caldirola, FordO'Connell, deOcaCaboBizet, Yaghjian, Medina, Sokolov, Hadad, ModifiedWithNumerics, Aguirregabiria, JohnsonHuEquation}, predominantly by reducing the equation for the particle's trajectory to a more familiar $2^{nd}$ order ODE, although this goes against the breadth of rigorous derivations for the AL/ALD equations \cite{Abraham, Lorentz1, Lorentz2, Dirac, ALD-Jackson, ALD-Rohrlich, Rohrlich2, BirnholtzHadarKol2013a, BirnholtzHadarKol2014a, BirnholtzHadar2013b, Gralla, Bopp, Kijowski, BambusiNoja, CaratiGalgani}.

Retaining the original equation, Dirac \cite{Dirac} suggested introducing an un-intuitive termination condition rather than the extra initial condition, requiring both the existence and value of the terminal acceleration\footnote{
Some have even further demanded the terminal acceleration to be zero \cite{Kosyakov};  others relaxed the condition to $e^{-t/\tau}|{\bf a}|\to0$ \cite{ModifiedWithNumerics}.
}.
However, even he realized this was a very contrived way of posing a physical question (backwards)
\footnote
{In his words,
``We now have a striking departure from the usual ideas of mechanics.
We must obtain solutions of our equations of motion for which the initial position and velocity of the electron are prescribes, together with its final acceleration, instead of solutions with all the initial conditions prescribed" \cite{Dirac}.
}
and an invitation for causality violation.
This condition also limits the scope of problems which can be solved to those where the acceleration approaches a finite constant at some (infinite or known finite) future time, and thus for many systems cannot be used.
For example, the classical system of an EM (Coulombic) two-body system loses energy to radiation while shrinking its orbit, causing the orbital frequency and the accelerations to grow boundlessly, rather than approach a finite limit.
Analogous systems have been treated successfully for general relativistic (GR) gravitational systems (\cite{HulseTaylor, Hulse, Taylor}, see also Sec. \ref{Gravity, Quantum Mechanics, and scales}), and Dirac's condition falls short of helping.

Behind an opposite approach \cite{Bopp, Kijowski, BambusiNoja, CaratiGalgani} lies the realization that the charge is not isolated and that the radiation does not arise from the particle alone; the complete system of course consists of both the charge and the EM field everywhere.
Realizing the missing information lies not in the particle's \emph{future} but in its \emph{past} and in its \emph{surroundings}, an alternative suggestion for initial data has been to supplement the particle's initial position and velocity by full Cauchy data for the entire field, everywhere.
This poses the problem at the opposite extreme: while in this approach no data is missing or guessed a-posteriori, too much information is required.
This is because the initial conditions for the particle and for the field are not independent, and are in fact related via the field singularity at the particle's position - with the dependence itself involving also velocity and acceleration (for example via the Li\'enard-Wiechert potential \cite{ALD-Jackson}) - making the problem over-constrained.

\sect {EFT with no Incoming Radiation}				%
From the perspective of Effective Field Theory (EFT), we wish to separate the charge degrees of freedom from those of the fields.
First we use the linearity of Maxwell's field equations to separate the charge and the field it itself generates from the background field.
The remaining system of the charge coupled to its own field is described with a simple Action formulation of a source-field radiating system; we thus employ the standard methods of dimensional reduction, field doubling, and then integrating out the field degrees of freedom (``Balayage" - for a full explanation of the method see \cite{BirnholtzHadarKol2013a}, for concise derivations of the AL/ALD radiation-reaction force see \cite{BirnholtzHadarKol2014a, BirnholtzHadar2013b}, and for EFT background \cite{CTP, GalleyEFT, GoldbergerRothstein1, GalleyNonConservative}).
We remain with an effective action describing the charge and the radiation-reaction on it, from which the Euler-Lagrange equation finally gives the equations of AL/ALD, with its high order time derivatives.

The key point lies in integrating out the radiation field: to do that, we solve the $2^{nd}$ order wave equation equation away from the charge, finding two solutions which describe incoming and outgoing radiation.
Upon plugging the solution back into the action to receive an effective action for the particle, we choose the solution of outgoing radiation, dropping the solution of incoming radiation, which amounts to the physical constraint of no incoming radiation.

This constraint of ``no incoming radiation" is precisely the desired supplementary initial condition.
It complements the initial conditions on particle's mechanical degrees of freedom (of which there are only position and velocity, as usual) with an initial condition on the radiation field at past infinity (often denoted $\scrim$ in GR terminology \cite{HawkingEllis}).
It guarantees the particle cannot ``drain" energy or angular momentum in from infinitely past and far away, but rather only radiates them outwards towards the future, which is physically desirable.

\sect {Benefits}				%

\subsection {Comparisons with previous approaches}		%

Comparing with the requirement for full Cauchy data, this condition is more lenient and does away with over-constraints, as it imposes a condition only on the radiation field on $\scrim$ rather than on the complete background+radiation field everywhere (for more on such separations, see \cite{DetweilerWhiting}).
Comparing with Dirac's termination condition, we first see that there is a class of problems which we are now able to pose and solve, but could not using his condition.
These include the aforementioned Coulombic 2-body inspiral and other systems which can not approach a finite terminal acceleration, but are governed by radiating away (without absorbing) energy and angular momentum.
Regarding the problems which did adhere to Dirac's condition of terminal acceleration, from our condition we learn that they may not draw energy from radiation at any time in their history (not only terminally so), and thus that any non-zero terminal acceleration must be the consequence of background fields.
In particular, systems where the terminal acceleration is non-zero must have a background field with unbounded spatial support - a physically unreasonable set-up.

A common practice regarding high derivatives in EFT is to treat them as small perturbations, first solving the low-order differential equations and only then adding the perturbative effects of the high derivatives \cite{WilsonKogut, Georgi, Polchinski}.
Similarly many authors have suggested replacing the ALD equation itself with $2^{nd}$ order ODE's, by effectively substituting for the high derivative term only the leading order \cite{Eliezer, LandauLifshitz} or a series expansion \cite{deOcaCaboBizet}.
These derivative expansions serve well to get rid of runaway solutions (by definition, because they are perturative), but have no regard for radiation absorption vs emission.
Thus on their own, they might still produce erroneous solutions; under the restriction of no incoming radiation, they may be a good practical method.

We thus see the (sometimes implicit) condition of no incoming radiation both expands the problem set we may look at, adds physical insight, and corrects erroneous behaviours.

We also note that the Balayage process of tracing the field outwards to infinity and then back to the particle, treated as a point from outside, avoids the questions of particle internals and finite-size effects, such as those of bare / negative mass.

\subsection {Further Implications}			%
Examining initial conditions can also help resolve the paradox regarding the energy emitted by a particle experiencing a constant non-zero acceleration ${\bf a}$.
In this case the work done by the AL force is (using \refeq{AL})
\be
P = {\bf v} \cdot \frac{2}{3} \frac{q^2}{c^3} \dot{\bf a} = 0,
\ee
while the Larmor formula predicts an average energy output of
\be
P = \frac{2}{3} \frac{q^2}{c^3} {\bf a}^2 > 0,
\ee
in apparent contradiction.
However, as the latter gives an only average over time, and as energy absorbed or radiated by the particle should only be defined asymptotically, a true comparison must refer to the time integrals of both formulae over the entire trajectory.
These two integrals are trivially related by integration by parts, and thus their difference is seen to reside entirely in the boundary terms, i.e. in the initial and terminal accelerations.
For acceleration over a finite time, this entirely suffices, as the boundary terms vanish.
This argument breaks for eternal acceleration - but in this situation, according to the Weak Equivalence Principle \cite{Equivalence} between acceleration and curvature, spacetime itself is no longer asymptotically flat (but rather Rindler space \cite{Rindler}), and thus the ``no incoming radiation" condition must be applied to those asymptotics, altering the definition of the radiation field and of the energy as viewed from infinity.

\begin{figure}[h]
\includegraphics[width=60mm]{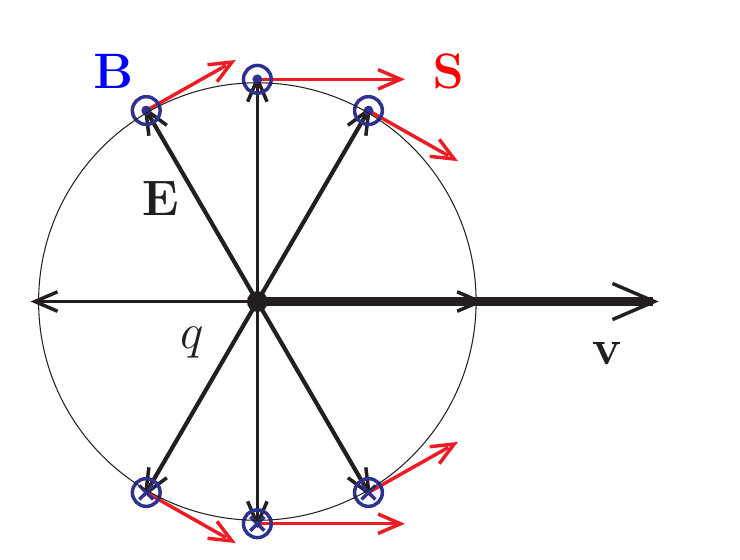}
\caption{The field of an electric charge $q$ moving at constant velocity ${\bf v}$. Shown are the electric (black) and magnetic (blue) field components, as well as the resulting Poynting vector (red). EM energy density follows the particle, but there is no radial energy flux towards or away from it.}
\label{fields_moving_charge}
\end{figure}

\begin{figure}[h]
\includegraphics[width=60mm]{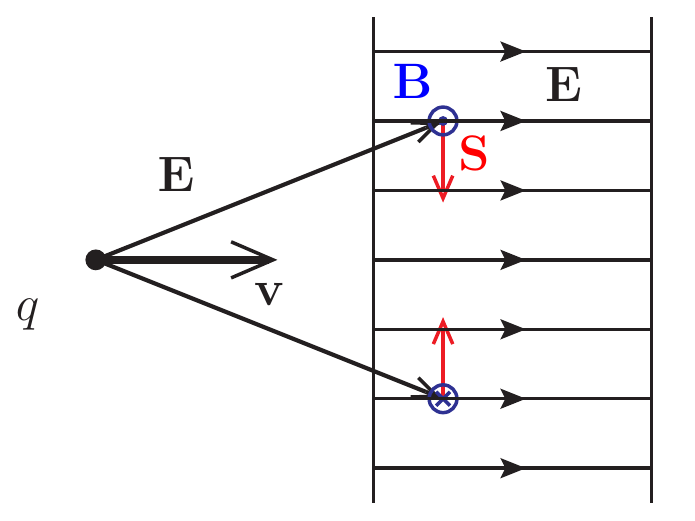}
\caption{The field of an electric charge $q$ approaching a region of constant electric field ${\bf E}$. The particle's field precedes it and penetrates the region with the constant electric field before the particle does. It causes energy flow and a change in EM fields within the region - changes which can radiate back towards the particle and affect it even before it enters the region itself, without breaking causality.}
\label{fields_moving_charge_penetrating}
\end{figure}

The paradox of pre-acceleration is usually posed thus: given an external electric field ``turned-on" at some point in time, the equation of motion for a charge (including the external force and the AL force) admits a solution involving exponentials of time which show that the particle accelerates before the field is turned on, in apparent violation of causality.
As an electric field cannot be turned on instantaneously, and as a time-varying field necessitates magnetic, current and/or radiation effects, we prefer the simpler and more realistic eternal electric field, confined to a region in space the particle enters in its trajectory.
A similar solution of the equation for such a background field shows the particle accelerates even before penetrating the domain of the external field.

Here again the surprise fades upon consideration of the incoming radiation from the infinite past.
\Refig{fields_moving_charge} shows the well-known fact that a charge moving with constant velocity in vacuum neither gives away or absorbs energy (the energy of the EM field rotates around the particle and follows its trajectory). 
\Refig{fields_moving_charge_penetrating} shows that the situation is different in the vicinity of an external electric field, where the motion of the particle creates an energy flow within the domain of the field, drawing energy from the domain boundaries inwards.
The particle's fields may also cause rearrangements of the charges sustaining the external field, who in turn affect the approaching particle.
This begins even before the particle itself penetrates the field's domain, because even before the particle penetrates, its own EM field already has; causality is preserved for the particle and the fields.

\sect{Conclusions: EM, QM \& Gravity}		%
\label{Gravity, Quantum Mechanics, and scales}		%
The problem of moving bodies emitting radiation, losing energy and angular momentum to it, and in turn experiencing a reaction force has also been studied extensively in the context of GR.
At the limit of slow velocities, the energy loss and radiation reaction are described by the Quadrupole formula \cite{GWEinstein}, which can be thought of as an analogue of the AL formula.
While AL describes dipole radiation (and thus involves 3 time derivatives), the Quadrupole formula involves 5 time derivatives, and is non-linear in the coordinates.

Introducing the physical properties of an electron in \refeq{AL}, we find the typical timescale to be of order $\sim10^{-24}s$, which is the time it takes light to cross the classical electron's radius $\sim10^{-15}m$.
The Quantum Mechanical (QM) nature of the electron becomes relevant at much longer space/time scales, i.e the Compton wavelength $\sim10^{-12}m$.
We thus expect classical EM radiation-reaction effects to be masked by QM behaviours (for detailed discussions see \cite{JohnsonHuEquation, JohnsonHu}).
The situation is reversed for gravitaional interactions, where the gravitational lengthscale is always at and above the Schwarzschild radius (at least $\sim10^3m$ for a compact object of stellar mass), many orders of magnitude above the scale for quantum gravity, expected at the Planck scale $\sim10^{-35}m$.
Hence the 2-body problem of an inspiraling orbit, for example, is a classical problem in GR and features prominently in searches for gravitational waves, while in EM it must be treated quantum mechanically, and has in fact historically led to the very development of QM \cite{Bohr}.
Discussions of classical radiation reaction are therefore much more natural to gravitational contexts than to EM.

However, aside from the quantum barrier relevant to one and not the other, the two problems are quite similar, and there is much mutual insight to be gained.
In particular, examinations of the asymptotics of spacetime, and the initial conditions at $\scrim$, have shed light on a century-old problem in EM.
Also, we hold that much can be learned about gravitational systems from studying the simpler corresponding problems in classical EM, where the EM case has the added benefits of linearity and less time derivatives (the GR Quadrupole formula has 5 time derivatives rather than ALD's 3).
Disregarding QM effects in such studies is legitimate when conclusions are desired eventually for gravitational systems, where QM is less relevant.\\

\begin{acknowledgments}						%
The author thanks B. Kol, A. Yarom, S. Hadar, B. Remez and P. Beniamini for helpful comments and discussions.
This research was supported by an ERC Advanced grant to Tsvi Piran and by the Israel Science Foundation grant no. 812/11.
It was part of the Einstein Research Project "Gravitation and High Energy Physics", which is funded by the Einstein Foundation Berlin.
\end{acknowledgments}





\begin{thebibliography}{99}


\bibitem{Abraham}
  M.~Abraham,
\VERBOSE{  ``Prinzipien der Dynamik des Elektrons,''}
  Annalen der Physik {\bf 10}, 105–179 (1903).

\bibitem{Lorentz1}
  H.~A.~Lorentz,
\VERBOSE{  ``Electromagnetic phenomena in a system moving with any velocity smaller than that of light,''}
  Proceedings of the Royal Netherlands Academy of Arts and Sciences {\bf 6}, 809–831 (1904).

\bibitem{Lorentz2}
  H.~A.~Lorentz,
  ``The Theory of Electrons and Its Applications to the Phenomena of Light and Radiant Heat,''
  Second Edition (Dover Publications, Inc., New York, 1952). The First edition appeared in 1909.

\bibitem{Dirac}
  P.~A.~M.~Dirac,
\VERBOSE{  ``Classical theory of radiating electrons,''}
  Proc.\ Roy.\ Soc.\ Lond.\ A {\bf 167}, 148 (1938).

\bibitem{ALD-Jackson}
  J.~D.~Jackson,
  ``Classical Electrodynamics,"
  Third Edition (Wiley, New York, 1998).

\bibitem{ALD-Rohrlich}
  F.~Rohrlich,
  ``Classical Charged Particles,"
  Third Edition (World Scientific Pub Co Inc, Singapore, 2007).

 \bibitem{Rohrlich2}
  F.~Rohrlich,
\VERBOSE{  ``The dynamics of a charged sphere and the electron,''}
  Am.\ J.\ Phys. {\bf 65} (11) p. 1051 (1997).



\bibitem{Plass}
  G.~N.~Plass,
\VERBOSE{  ``Classical Electrodynamic Equations of Motion with Radiative Reaction,''}
  Rev.\ Mod.\ Phys.\ {\bf 33}, 37 (1961).

\bibitem{Driver}
  R.~D.~Driver,
\VERBOSE{  ``A Two-Body Problem of Classical Electrodynamics: the One-Dimensional Case,"}
  Ann.\ of Phys.\ {\bf 21}, 122-142 (1963).
  R.~D.~Driver and M.~J.~Norris,
\VERBOSE{  ``Note on Uniqueness for a One-Dimensional Two-Body Problem of Classical Electrodynamics,"}
  Ann.\ of Phys.\ {\bf 42}, 347-351 (1967).

\bibitem{WheelerFeynman}
  J.~A.~Wheeler abd R.~P.~Feynman,
\VERBOSE{  ``Classical Electrodynamics in Terms of Direct Interparticle Action,"}
  Rev.\ of Mod.\ Phys.\ {\bf 21}, 425–433 (1949).

\bibitem{Murdock}
  J.~A.~Murdock,
\VERBOSE{  ``On the Well-Posed Two-Body Problem in Electrodynamics and Special Relativity,"}
  Ann.\ of Phys.\ {\bf 84}, 432-439 (1974).

\bibitem{Spohn}
  H.~Spohn,
\VERBOSE{  ``The Critical manifold of the Lorentz-Dirac equation,''}
  Europhys.\ Lett.\  {\bf 50}, 287 (2000)
  [physics/9911027].

\bibitem{HerasDebate}
  J.~A.~Heras,
\VERBOSE{  ``Preacceleration without radiation: The nonexistence of preradiation phenomenon,"}
  Am.\ J.\ Phys.\ {\bf 74}, 1025–1030 (2006).
  J.~D.~Jackson,
\VERBOSE{  ``Comment on ``Preacceleration without radiation: The nonexistence of preradiation phenomenon," by J. A. Heras [Am. J. Phys.74 (11), 1025–1030 (2006)]",}
  Am.\ J.\ Phys.\ {\bf 75}, 844 (2007).
  V.~Hnizdo,
\VERBOSE{  ``Comment on `Preacceleration without radiation: The nonexistence of preradiation phenomenon,’ by J. A. Heras [Am. J. Phys. 74, 1025–1030 (2006)],"}
  Am.\ J.\ Phys.\ {\bf 75}, 845-846 (2007).
  J.~A.~Heras,
\VERBOSE{  ``Reply to ``Comment(s) on ‘Preacceleration without radiation: The nonexistence of preradiation phenomenon,'" by J. D. Jackson [Am. J. Phys.75 (9), 844–845 (2007)] and V. Hnizdo [Am. J. Phys.75 (9), 845–846 (2007)]",}
  Am.\ J.\ Phys.\ {\bf 75}, 847 (2007).


\bibitem{Laue}
  M.~von Laue,
  ``Das Relativit\"atsprinzip,"
  (Vieweg, Braunschweig, 1911).

\bibitem{Fermi}
  E.~Fermi,
\VERBOSE{  ``\"Uber einen Widerspruch zwischen der elektrodynamischen und relativistischen Theorie der elektromagnetischen Masse"
  [``Concerning a Contradiction between the Electrodynamic and Relativistic Theory of Electromagnetic Mass"],}
  Physikalische Zeitschrift 23: 340–344' (1922).

\bibitem{Rohrlich3}
  F.~Rohrlich,
\VERBOSE{  ``Self-Energy and Stability of the Classical Electron",}
  Am.\ J.\ Phys. {\bf 28} (7) 639-643 (1960).

\bibitem{SchwingerEM}
  J.~Schwinger,
\VERBOSE{  ``Electromagnetic mass revisited,"}
  Foundations of Physics {\bf 13}  (3) : 373–383 (1983).

\bibitem{Kosyakov}
  B.~P.~Kosyakov,
  ``Introduction to the classical theory of particles and fields,''
  Berlin, Germany: Springer (2007).

\bibitem{Steane}
  A.~M.~Steane,
\VERBOSE{  ``Reduced-order Abraham-Lorentz-Dirac equation and the consistency of classical electromagnetism,'' (2014)}
  [arXiv:1402.1106 [physics.class-ph]].

\bibitem{Rohrlich1}
  F.~Rohrlich,
\VERBOSE{  ``Classical selfforce,''}
  Phys.\ Rev.\ D {\bf 60}, 084017 (1999).

\bibitem{Parrott}
  S.~Parrott,
\VERBOSE{  ``Comment on Phys. Rev. D 60 084017 `classical self-force' by F. Rohrlich,''}
  gr-qc/0502029.



\bibitem{Knoll}
  Y.~Knoll,
\VERBOSE{  ``What went wrong with physics? The classical self force problem revisited, with radical implications to all of physics,'' (2014)}
  [arXiv:1201.5281 [physics.gen-ph]].

\bibitem{Eliezer}
  C.~J.~Eliezer,
\VERBOSE{  ``On the Classical Theory of Particles,"}
  Proc.\ Royal.\ Soc.\ Lond.\ A {\bf 194}, 543 (1948).

\bibitem{LandauLifshitz}
  L.~D.~Landau and E.~M.~Lifshitz,
  ``The Classical Theory of Fields,"
  Second Edition (Addison-Wesley, Cambridge, 1951).

\bibitem{MoPapas}
  Tse Chin Mo and C.~H.~Papas,
\VERBOSE{  ``New Equation of Motion for Classical Charged Particles,"}
  Phys.\ Rev.\ D {\bf 4}, 3566 (1971).

\bibitem{Caldirola}
  P.~Caldirola,
\VERBOSE{  ``A Relativistic Theory of the Classical Electron,"}
  Rivista del Nuovo Cimento {\bf 2}, {\bf 13} p.1-49 (1979).

\bibitem{FordO'Connell}
  G.~W.~Ford and R.~F.~O'Connell,
  Phys.\ Lett.\ A {\bf 157}, 217 (1991).
  G.~W.~Ford and R.~F.~O'Connell,
  Phys.\ Rev.\ A {\bf 44}, 6386 (1991).
  G.~W.~Ford and R.~F.~O'Connell,
  Phys.\ Lett.\ A {\bf 174}, 182 (1993).
  G.~W.~Ford and R.~F.~O'Connell,
\VERBOSE{  ``Alternative equations of motion for the radiating electron,"}
  Applied Physics B {\bf 60}, 301-302 (1995).

\bibitem{deOcaCaboBizet}
  A.~C.~M.~de Oca and N.~G.~Cabo-Bizet,
\VERBOSE{  ``Newton-like equations for a radiating particle,''}
  arXiv:1309.3213 [gr-qc] (2013).

\bibitem{Yaghjian}
  A.~D.~Yaghjian,
  ``Relativistic Dynamics of a Charged Sphere: Updating the Lorentz–Abraham Model,"
  Lecture Notes in Physics 686,
  Second Edition (Springer, New York, 2006).

 \bibitem{Medina}
  R.~Medina,
\VERBOSE{  ``Radiation reaction of a classical quasi-rigid extended particle,''}
  J.\ Phys.\ A {\bf A39}, 3801 (2006)
  [physics/0508031].

\bibitem{Sokolov} 
  I.~V.~Sokolov et al., 
  Journ.\ Exp.\ Theor.\ Phys.\ {\bf 109}, 207 (2009).
  I.~V.~Sokolov et al.,
\VERBOSE{  ``Dynamics of emitting electrons in strong laser fields,"}
  Phys.\ Plasmas {\bf 16}, 093115 (2009).
  I.~V.~Sokolov, J.~A.~Nees, V.~P.~Yanovsky, N.~M.~Naumova and G.~A.~Mourou,
\VERBOSE{  ``Emission and its back-reaction accompanying electron motion in relativistically strong and QED-strong pulsed laser fields,''}
  Phys.\ Rev.\ E {\bf 81}, 036412 (2010)
  [arXiv:1003.0806 [physics.plasm-ph]].

\bibitem{Hadad}
  Y.~Hadad, L.~Labun, J.~Rafelski, N.~Elkina, C.~Klier and H.~Ruhl,
\VERBOSE{  ``Effects of Radiation-Reaction in Relativistic Laser Acceleration,''}
  Phys.\ Rev.\ D {\bf 82}, 096012 (2010)
  [arXiv:1005.3980 [hep-ph]].

 \bibitem{ModifiedWithNumerics}
  G.~G.~Alcaine, F.~J.~Llanes-Estrada,
\VERBOSE{  ``Radiation reaction on a classical charged particle: a modified form of the equation of motion,''}
  arXiv:1211.5486v2 [physics.class-ph] (2013).

 \bibitem{Aguirregabiria}
  J.~M.~Aguirregabiria, J.~Llosa, and A.~Molina,
\VERBOSE{  ``Motion of a classical charged particle,''}
  Phys.\ Rev.\ D {\bf 73}, 125015  (2006).

\bibitem{JohnsonHuEquation}
  P.~R.~Johnson and B.~L.~Hu,
\VERBOSE{  ``Uniformly accelerated charge in a quantum field: From radiation reaction to Unruh effect,''}
  Found.\ Phys.\  {\bf 35} (2005) 1117
  [gr-qc/0501029].

\bibitem{Gralla} 
  S.~E.~Gralla, A.~I.~Harte and R.~M.~Wald,
\VERBOSE{  ``A Rigorous Derivation of Electromagnetic Self-force,''}
  Phys.\ Rev.\ D {\bf 80}, 024031 (2009)
  [arXiv:0905.2391 [gr-qc]].


\bibitem{BirnholtzHadarKol2013a}
  O.~Birnholtz, S.~Hadar and B.~Kol,
\VERBOSE{  ``Theory of post-Newtonian radiation and reaction,''}
  Phys.\ Rev.\ D {\bf 88}, 104037 (2013)
  [arXiv:1305.6930 [hep-th]].

\bibitem{BirnholtzHadarKol2014a}
  O.~Birnholtz, S.~Hadar and B.~Kol,
\VERBOSE{  ``Radiation reaction at the level of the action,''}
  Int.\ J.\ Mod.\ Phys.\ A {\bf 29}, no. 24, 1450132 (2014)
  [arXiv:1402.2610 [hep-th]].

\bibitem{BirnholtzHadar2013b}
  O.~Birnholtz and S.~Hadar,
\VERBOSE{  ``An action for reaction in general dimension,''}
  Phys.\ Rev.\ D {\bf 89}, 045003 (2014)
  [arXiv:1311.3196 [hep-th]].

\bibitem{Bopp}
  F.~Bopp,
\VERBOSE{  ``Lineare Theorie des Elektrons. II,''}
  Ann.\ Der.\ Phys.\  {\bf 42}, 573-608 (1943).

\bibitem{Kijowski}
  J.~Kijowski,
\VERBOSE{  ``Electrodynamics of moving particles,''}
  GRG  {\bf 26}, 167-201 (1994).

\bibitem{BambusiNoja}
  D.~Bambusi and D.~Noja,
\VERBOSE{  ``On classical electrodynamics of point particles and mass renormalization: some preliminary results,''}
  Lett.\ in Math.\ Phys.\ {\bf 37} (4), 449-460 (1996).

\bibitem{CaratiGalgani}
  A.~Carati and L.~Galgani,
\VERBOSE{  ``Recent progress on the Abraham-Lorentz-Dirac equation,''}
  in New Perspective in the physics of mesoscopic system,
  S.~ De Martino et al. eds.,
  World Scientific (Singapore, 1997).

\bibitem{HulseTaylor}
  R.~A.~Hulse and J.~H.~Taylor,
\VERBOSE{  ``Discovery of a pulsar in a binary system,''}
  Astrophys.\ J.\  {\bf 195}, L51 (1975).

\bibitem{Hulse}
  R.~A.~Hulse,
\VERBOSE{  ``The discovery of the binary pulsar,''}
  Rev.\ Mod.\ Phys.\  {\bf 66}, 699 (1994).

\bibitem{Taylor}
  J.~H.~Taylor,
\VERBOSE{  ``Binary pulsars and relativistic gravity,''}
  Rev.\ Mod.\ Phys.\  {\bf 66}, 711 (1994).


\bibitem{CTP}
  J.~S.~Schwinger,
\VERBOSE{  ``Brownian motion of a quantum oscillator,''}
  J.\ Math.\ Phys.\  {\bf 2}, 407 (1961).
  K.~T.~Mahanthappa,
\VERBOSE{  ``Multiple production of photons in quantum electrodynamics,''}
  Phys.\ Rev.\  {\bf 126}, 329 (1962).
  L.~V.~Keldysh,
\VERBOSE{  ``Diagram technique for nonequilibrium processes,''}
  Zh.\ Eksp.\ Teor.\ Fiz.\  {\bf 47}, 1515 (1964)
  [Sov.\ Phys.\ JETP {\bf 20}, 1018 (1965)].

\bibitem{GalleyEFT}
  C.~R.~Galley and B.~L.~Hu,
\VERBOSE{  ``Self-force with a stochastic component from radiation reaction of a scalar charge moving in curved spacetime,''}
  Phys.\ Rev.\ D {\bf 72}, 084023 (2005)
  [gr-qc/0505085].
  C.~R.~Galley and M.~Tiglio,
\VERBOSE{  ``Radiation reaction and gravitational waves in the effective field theory approach,''}
  Phys.\ Rev.\ D {\bf 79}, 124027 (2009)
  [arXiv:0903.1122 [gr-qc]].

\bibitem{GoldbergerRothstein1}
  W.~D.~Goldberger and I.~Z.~Rothstein,
\VERBOSE{ ``An effective field theory of gravity for extended objects,''}
  Phys.\ Rev.\  D {\bf 73}, 104029 (2006).


\bibitem{GalleyNonConservative}
  C.~R.~Galley,
\VERBOSE{  ``The classical mechanics of non-conservative systems,''}
  Phys.\ Rev.\ Lett.\  {\bf 110}, 174301 (2013)
  [arXiv:1210.2745 [gr-qc]].


\bibitem{HawkingEllis}
  S.~W.~Hawking and G.~F.~R.~Ellis,
  ``The Large scale structure of space-time,''
  Cambridge University Press, Cambridge, 1973.

\bibitem{DetweilerWhiting}
  S.~L.~Detweiler and B.~F.~Whiting,
\VERBOSE{  ``Selfforce via a Green's function decomposition,''}
  Phys.\ Rev.\ D {\bf 67}, 024025 (2003)
  [gr-qc/0202086].

\bibitem{WilsonKogut}
  K.~G.~Wilson and J.~B.~Kogut,
\VERBOSE{  ``The Renormalization group and the epsilon expansion,''}
  Phys.\ Rept.\  {\bf 12}, 75 (1974).

\bibitem{Polchinski}
  J.~Polchinski,
\VERBOSE{  ``Renormalization and Effective Lagrangians,''}
  Nucl.\ Phys.\ B {\bf 231}, 269 (1984).

\bibitem{Georgi}
  H.~Georgi,
\VERBOSE{  ``Effective field theory,''}
  Ann.\ Rev.\ Nucl.\ Part.\ Sci.\  {\bf 43}, 209 (1993).

\bibitem{Equivalence}
  M.~P.~Haugen and C.~L\"{a}mmerzahl,
  ``Principles of Equivalence: Their Role in Gravitation Physics and Experiments that Test Them"
  (Springer, New York, 2001)
  [arXiv:gr-qc/0103067].

\bibitem{Rindler}
  W.~Rindler,
\VERBOSE{  ``Kruskal Space and the Uniformly Accelerated Frame,''}
  Am.\ J.\ Phys.\  {\bf 34}, 1174 (1966).

\bibitem{GWEinstein}
  A.~Einstein,
\VERBOSE{  ``\"{U}ber Gravitationswellen''}
  Sitzungsberichte der K\"{o}niglich Preussischen Akademie der Wissenschaften Berlin (1918), 154-167.

\bibitem{JohnsonHu}
  B.~L.~Hu and P.~R.~Johnson,
\VERBOSE{  ``Beyond Unruh effect: Nonequilibrium quantum dynamics of moving charges,''}
  quant-ph/0012132.
  P.~R.~Johnson and B.~L.~Hu,
\VERBOSE{  ``Worldline influence functional: Abraham-Lorentz-Dirac-Langevin equation from QED,''}
  quant-ph/0012135.
  P.~R.~Johnson and B.~L.~Hu,
\VERBOSE{  ``Stochastic theory of relativistic particles moving in a quantum field. 1. Influence functional and Langevin equation,''}
  quant-ph/0012137.
  P.~R.~Johnson and B.~L.~Hu,
\VERBOSE{  ``Stochastic theory of relativistic particles moving in a quantum field. 2. Scalar Abraham-Lorentz-Dirac-Langevin equation, radiation reaction and vacuum fluctuations,''}
  Phys.\ Rev.\ D {\bf 65}, 065015 (2002)
  [quant-ph/0101001].

\bibitem{Bohr}
  N.~Bohr,
\VERBOSE{  ``I. On the constitution of atoms and molecules''}
  Philosophical Magazine {\bf 26} (151), 1-25 (1913).

\end{thebibliography}
\end{document}